\documentclass[pra,twocolumn,showkeys,showpacs,preprintnumbers,amsmath,amssymb]{revtex4}
\usepackage{dsfont}     
\usepackage{amsmath}    
\usepackage{amsthm}     
\usepackage{amssymb}    

\usepackage{color}      
\usepackage{graphicx} 	

\newtheorem{theorem}{Theorem} 
\newtheorem{lemma}[theorem]{Lemma}

\newcommand{\bra}[1]{\left\langle #1 \right\vert}   
\newcommand{\ket}[1]{\left\vert#1\right\rangle}     
\newcommand{\braket}[2]{\left\langle #1\vert#2\right\rangle}    
\newcommand{\mel}[3]{\left\langle #1 \left\vert#2\right\vert#3\right\rangle} 
\newcommand{\abs}[1]{\left\vert#1\right\vert}   

\begin{document}

\title{On the Geometric Measures of Entanglement}

\author{K. Uyan\i k}
 \email{e120175@metu.edu.tr}
\author{S. Turgut}
 \email{sturgut@metu.edu.tr}
\affiliation{Department of Physics, Middle East Technical University\\
06531, Ankara, TURKEY}

\date{\today}

\begin{abstract}
The geometric measure of entanglement, which expresses the minimum
distance to product states, has been generalized to distances to
sets that remain invariant under the stochastic reducibility
relation. For each such set, an associated entanglement monotone
can be defined. The explicit analytical forms of these measures
are obtained for bipartite entangled states. Moreover, the three
qubit case is discussed and argued that the distance to the W
states is a new monotone.
\end{abstract}



\maketitle

\section{Introduction}
\label{sec:Introduction}

Entanglement is a profound phenomenon which intrigued scientists with both
its rich mathematical structure and its deep philosophical implications.
In the last two decades it has also come into prominence as a key resource
in quantum communication and quantum computation\cite{Horodecki09}.
Quantifying entanglement is one of the main problems of quantum
information theory. Although for bipartite pure states entanglement
measures are well established, quantification of mixed state and
multipartite pure state entanglement still contain many unresolved
issues\cite{Horodecki09,Plenio07,Bennett00}. For multipartite states, not
only the amount but also the flavor of entanglement becomes pertinent. For
example, for three qubits there are two distinct flavors of genuine
tripartite entanglement represented by the Greenberger-Horne-Zeilinger
(GHZ) and W states which can never be converted into each
other\cite{Dur00}. The essential property that is not present or lacking a
proper analogue in these settings is the Schmidt decomposition. Still, a
number of important applications require a proper measure for systems
beyond the bipartite pure setting and the challenge continues.

A considerable number of multipartite entanglement measures are
generalizations of bipartite measures. Geometric measure of
entanglement\cite{Wei03} is one of them. It is introduced by
Shimony \cite{Shimony95} and generalized to multipartite states by
Barnum and Linden\cite{Barnum01}. Wei and Goldbart extended it to
multipartite mixed states by employing the convex roof
construction\cite{Wei03}.

For a pure state $\ket{\psi}$, the geometric distance is defined
as the minimum distance between $\ket{\psi}$ and the set of
product states\cite{Wei03}, i.e.,
\begin{equation}
  d(\psi,S)=\inf_{\ket{\phi}\in S}\left\Vert \ket{\psi} -\ket{\phi}  \right\Vert~~,
\label{eq:d_defined}
\end{equation}
where $S$ denotes the set of normalized product states. Even
though the distance has a clear geometric meaning, an entanglement
monotone\cite{Vidal00}, i.e., a quantity which never increase on
the average under local quantum operations assisted with classical
communication (LOCC) is more useful in applications. The quantity
\begin{equation}
  E(\psi,S) = 1- \sup_{\ket{\phi}\in S}\abs{\braket{\psi}{\phi}}^2~~,
\label{eq:E_defined}
\end{equation}
satisfies this property\cite{Wei03}. Moreover, it is related to
the geometric distance by a monotone increasing function as
\begin{equation}
  d(\psi,S)=\sqrt{2\left( 1 - \sqrt{1-E(\psi,S)}\right)}\quad.
\end{equation}

The geometric measure is used in a number of applications for
expressing the degree of entanglement. It is related to the
Groverian measure of entanglement\cite{Biham01}, a quantity which
measures the degree of success in Grover's search algorithm\cite{Grover97}
 as a function of the initial entangled state used.
It is also involved, along with some other distancelike measures,
in an upper bound expression on the maximum number of multipartite
states that can be discriminated by LOCC\cite{Hayashi06}.
Recently, it is shown that for measurement-based quantum
computing, too much entanglement in the initial state, as measured
by the geometric measure, is detrimental for the quantum speedup
gained over classical algorithms\cite{Gross09}. Finally, the
geometric measure has also found applications in many-body
physics\cite{Wei05,Orus08}.

It is interesting to investigate the distance of the state
$\ket{\psi}$ to more general sets $S$ other than the product
states for the purpose of generalizing the geometric measure. In
this way, it is possible to quantify the flavor of entanglement
more directly; i.e., the degree of the difference of the
entanglement in state $\ket{\psi}$ from those of the states in $S$
can be measured by $E(\psi,S)$. In addition to this, when the
state $\ket{\psi}$ is desired to be approximated by states in $S$,
the maximum achievable fidelity is deficient by $E(\psi,S)$ from
the absolute maximum $1$. For example, there might be situations
where the state $\ket{\psi}$ is required in a particular
quantum communication or computation task; but it may be too
difficult to establish this state between distant parties.
Instead, states in the set $S$ could be easily established with
little cost. For such cases, smallness of the deficiency
$E(\psi,S)$ can be used as a measure of replaceability of
$\ket{\psi}$ by some state in $S$. An experiment is discussed in
Ref. \onlinecite{Walther05} where GHZ class states are used for
obtaining approximate W states.

It will be shown that, if the set $S$ remains invariant under the
stochastic reducibility relation of Ref.~\onlinecite{Bennett00},
then the function $E(\psi,S)$ is also a monotone. In this case the
distance $d$ and the monotone $E$ satisfy properties similar to
those of the geometric measure of Wei and Goldbart. Hence, there
appears to be a number of different \emph{geometric measures}; the
case for which $S$ is the set of product states gives just one of
those measures. These measures also include previous
generalizations\cite{Barnum01,Blasone08}, but the current
generalization appears to exhaust all such pure-state measures
that are based on the geometric distance to a set. The purpose of
this article is to investigate these general geometric measures.
It may be the case that some of these measures might be useful in
the sense that they forbid some entanglement transformations
allowed by all other known monotones. Indeed, it appears that the
distance to the W class states of three qubits is an example of
this.

The organization of the article is as follows. In section
\ref{sec:DefinitionsAndMonotonicity}, the precise definition of
the key property of the sets $S$ is presented and the monotonicity
of $E(\psi,S)$ is proven. Section \ref{sec:Bipartite} contains the
computation of all of the geometric measures for the bipartite
entangled states. After that, the three qubit case is investigated
and one of the monotones is shown to be a new monotone in Section
\ref{sec:ThreeQubits}. Finally a brief summary is given in section
\ref{sec:Conclusion}.

\section{Definitions and Proof of Monotonicity}
\label{sec:DefinitionsAndMonotonicity}

In the following, $p$-partite entangled pure states between $p$
distant particles will be considered. Throughout this article we
will be interested in pure states only. Let us briefly recall the
following definitions. The entangled state $\ket{\phi}$ is said to
be \emph{stochastically reducible} to $\ket{\phi^\prime}$, if,
after starting with the initial state $\ket{\phi}$, it is possible
to obtain $\ket{\phi^\prime}$ with non-zero probability by LOCC
operations\cite{Bennett00}. This is equivalent to the existence of
local operators $A_i$ such that
\begin{equation}
  \ket{\phi^\prime}=N (A_1\otimes A_2\otimes \cdots\otimes A_p) \ket{\phi}
\label{eq:stochastic_reducibilty}
\end{equation}
for some normalization constant $N$. We say that $\ket{\phi}$ and
$\ket{\phi^\prime}$ are \emph{SLOCC equivalent} if they are
stochastically reducible to each other. That statement is
equivalent to the existence of invertible local operators $A_i$
such that Eq.~(\ref{eq:stochastic_reducibilty}) is satisfied
\cite{Dur00}. The equivalence classes obtained from this
equivalence relation are called SLOCC classes. If $\ket{\phi}$ is
stochastically reducible to $\ket{\phi^\prime}$, then all states
in the SLOCC class of $\ket{\phi}$ is stochastically reducible to
any state in the class of $\ket{\phi^\prime}$. In other words,
stochastic reducibility is also a relation between SLOCC classes.

Let $S$ be a set of states. If for any state $\ket{\phi}$ in $S$,
all states $\ket{\phi^\prime}$ that are stochastically reducible
from $\ket{\phi}$ are in $S$, then we will say that $S$ is a
\emph{stochastically invariant} (SI) or \emph{SLOCC invariant}
set. Obviously, if $\ket{\phi}$ is in the SI set $S$, then the
whole of the SLOCC class of $\ket{\phi}$ is a subset of $S$. This
means that SI sets are unions of SLOCC classes. Moreover, if $S$
contains a particular SLOCC class, then all classes that are
stochastically reducible from this class are also contained in
$S$. In particular, this implies that $S$ contains (the class of)
all product states. The central result of this article is the
following theorem.

\begin{theorem}
If $S$ is a SI set of normalized states, then the function
$E(\psi,S)$ of the normalized states $\ket{\psi}$ is an
entanglement monotone.
\end{theorem}

The requirements that $S$ contain only normalized states and
$\ket{\psi}$ is normalized are imposed for utilizing
Eq.~(\ref{eq:E_defined}) as a definition of $E$. Before proving
this theorem, we need the following lemma.

\begin{lemma}
\label{FirstLemma}
Let $\ket{\alpha},\ket{\beta_1},\ldots,\ket{\beta_n}$ be vectors in a Hilbert space.
Then, the operator inequality
\begin{equation}
  \ket{\alpha}\bra{\alpha} \leq \sum_{i=1}^n\ket{\beta_i}\bra{\beta_i}
\label{eq:Lemma_statement1}
\end{equation}
is satisfied  if and only if there exists numbers $c_i$ such that
\begin{equation}
  \ket{\alpha} = \sum_{i=1}^n c_i\ket{\beta_i} \quad\textrm{and}\quad  \sum_{i=1}^n \abs{c_i}^2 \leq 1~~.
\label{eq:Lemma_statement2}
\end{equation}
\end{lemma}

\textit{Proof:} We first show the necessity. Suppose that
Eq.~(\ref{eq:Lemma_statement1}) is satisfied. Then, $B=\sum_i
\ket{\beta_i}\bra{\beta_i}-\ket{\alpha}\bra{\alpha}$ is positive
semidefinite. Let $m$ be the number of non-zero eigenvalues of
$B$. By using the spectral decomposition of $B$, we can find $m$
vectors $\ket{\gamma_1},\ldots,\ket{\gamma_m}$ such that
$B=\sum_i\ket{\gamma_i}\bra{\gamma_i}$. Hence, we have the
following relation
\begin{equation}
  \sum_{i=1}^m  \ket{\gamma_i}\bra{\gamma_i} + \ket{\alpha}\bra{\alpha}=\sum_{i=1}^n\ket{\beta_i}\bra{\beta_i}
\end{equation}
Then, by the Schr\"odinger-GHJW theorem\cite{Hughston93}, there is
a $d\times d$ unitary matrix $U$ (where $d=\max(m+1,n)$), such
that
\begin{eqnarray}
  \ket{\gamma_j} &=& \sum_i U_{j,i} \ket{\beta_i}\quad (j=1,\ldots,m)~,\\
  \ket{\alpha}   &=& \sum_i U_{m+1,i} \ket{\beta_i}\quad.
\end{eqnarray}
Hence, we take $c_i=U_{m+1,i}$. Moreover,
\begin{equation}
\sum_{i=1}^n \abs{c_i}^2 = \sum_{i=1}^n \abs{U_{m+1,i}}^2 \leq \sum_{i=1}^d \abs{U_{m+1,i}}^2=1\quad.
\end{equation}
Therefore, (\ref{eq:Lemma_statement2}) is satisfied.

Now, for proving the sufficiency part, suppose that
(\ref{eq:Lemma_statement2}) is satisfied. Let $\ket{\gamma}$ be
any arbitrary vector. Then,
\begin{eqnarray}
 \braket{\gamma}{\alpha}\braket{\alpha}{\gamma} &=& \abs{ \braket{\gamma}{\alpha} }^2 \\
      &=& \abs{\sum_i c_i \braket{\gamma}{\beta_i}}^2  \label{eq:Schwarz1}\\
      &\leq& \left(  \sum_i \abs{c_i}^2  \right) \left(  \sum_i\abs{\braket{\gamma}{\beta_i}}^2  \right)  \label{eq:Schwarz2}\\
      &\leq& \sum_i \abs{\braket{\gamma}{\beta_i}}^2 = \sum_i \braket{\gamma}{\beta_i}\braket{\beta_i}{\gamma}~.
\end{eqnarray}
Here, Schwarz inequality is used in passing from
(\ref{eq:Schwarz1}) to (\ref{eq:Schwarz2}). Since the last
inequality is valid for all vectors $\ket{\gamma}$, then the
associated inequality for operators, i.e.,
(\ref{eq:Lemma_statement1}) is satisfied. This completes the proof
of the equivalence of (\ref{eq:Lemma_statement1}) and
(\ref{eq:Lemma_statement2}). $\Box$

Now, we can start with the proof of the theorem. It should be
shown that, if by LOCC, $\ket{\psi}$ is transformed to states
$\ket{\psi_i}$ with probability $p_i$, then
\begin{equation}
  E(\psi,S) \geq \sum_i p_i E(\psi_i,S)~~.
  \label{eq:monotone_ineq}
\end{equation}
For this purpose, it is enough to prove this inequality for the
local operations carried out by a single party only. Hence,
without loss of generality, it will be assumed that the first
party is carrying out a measurement. Let $M_i$ be the local
measurement operators associated with this operation. They satisfy
$\sum_i M_i^\dagger M_i =\mathds{1}_1$ where $\mathds{1}_i$ is
used for denoting the identity operator acting on the state space
of the $i$th party's particle. Hence, we have
\begin{eqnarray}
  p_i &=& \bra{\psi}(M_i^\dagger M_i \otimes \mathds{1}_2 \otimes\cdots  \otimes\mathds{1}_p)\ket{\psi}~~,\\
  \ket{\psi_i} &=&\frac{1}{\sqrt{p_i}}(M_i \otimes \mathds{1}_2 \otimes\cdots  \otimes\mathds{1}_p)\ket{\psi}~~.
\end{eqnarray}
Let $P_i=M_i^\dagger M_i$ and $U_i$ be the appropriate unitary
operators satisfying $M_i=U_i\sqrt{P_i}$, whose existence is
guaranteed by the polar decomposition of operators. For
simplifying the notation, boldface letters will be used for
denoting the 1st particle's local operators as an operator acting
on the whole state space, i.e., $\mathbf{P}_i = P_i
\otimes\mathds{1}_2\otimes\cdots\otimes\mathds{1}_p$ etc.

Let $\ket{\phi}$ be an arbitrary vector in $S$; it will be used in the
maximization of the right-hand side of Eq.~(\ref{eq:E_defined}). Let
\begin{equation}
    n_i = \bra{\phi} \mathbf{P}_i \ket{\phi}
\end{equation}
and note that these are non-negative numbers having the sum
$\sum_i n_i =1$. Let us define the vectors $\ket{\phi_i}$ in $S$
as
\begin{equation}
  \ket{\phi_i} = \left\{
    \begin{array}{ll} \
       n_i^{-1/2}\sqrt{\mathbf{P}_i}\ket{\phi} & \textrm{if}~n_i\neq0 \\
       \ket{\phi}   & \textrm{if}~n_i=0
    \end{array}
    \right.
    ~~~.
\end{equation}
Note that each vector $\ket{\phi_i}$ is normalized and
stochastically reducible from $\ket{\phi}$.
Therefore all of them are in $S$. For the case $n_i=0$, the value
of $\ket{\phi_i}$ is unimportant; it has just been assigned to a
vector known to exist in $S$.

First note that
\begin{eqnarray}
  \ket{\phi}=\sum_i \sqrt{n_i}  \sqrt{\mathbf{P}_i}\ket{\phi_i}\quad.
\end{eqnarray}
and $\sum_i n_i=1$. Hence, by applying the lemma, we see that
\begin{equation}
  \ket{\phi}\bra{\phi} \leq \sum_i \sqrt{\mathbf{P}_i}\ket{\phi_i} \bra{\phi_i} \sqrt{\mathbf{P}_i}
\end{equation}
holds as an operator inequality. If the expectation value in the
state $\ket{\psi}$ is taken, we get
\begin{eqnarray}
  \abs{ \braket{\phi}{\psi} }^2 & \leq &
      \sum_i \abs{ \mel{\phi_i}{ \sqrt{\mathbf{P}_i} }{\psi} }^2 \\
      &=& \sum_i p_i \abs{ \mel{ \phi_i}{\mathbf{U}_i^\dagger}{\psi_i} }^2 \\
      &=& \sum_i p_i \abs{ \braket{ \phi_i^\prime}{\psi_i}}^2
\end{eqnarray}
where $\ket{\phi^\prime_i}=\mathbf{U}_i\ket{\phi_i}$, which are
also in $S$. Using the obvious fact that a particular value of a
function is smaller than its maximum, i.e.,
$\abs{\braket{\phi_i^\prime}{\psi_i}}^2 \leq 1-E(\psi_i,S)$, we
get
\begin{equation}
   \abs{ \braket{\phi}{\psi} }^2 \leq \sum_i p_i (1-E(\psi_i,S))\quad.
\end{equation}
If the left-hand side is maximized over $\ket{\phi}$, then the
inequality (\ref{eq:monotone_ineq}) is obtained. Finally, we note
that $S$ contains product states and hence $E(\psi,S)=0$ vanishes
if $\ket{\psi}$ is a product state. This completes the proof of
the monotonicity of $E(\psi,S)$. $\Box$

Let us make a few remarks on the dependence of the measure $E$ in
Eq.~(\ref{eq:E_defined}) on the set $S$. It can be readily
observed that for any arbitrary set $S$, the topological closure
$\overline{S}$ yields the same measure, i.e.,
$E(\psi,S)=E(\psi,\overline{S})$. In addition to this, two
different sets yield the same measure $E$ (and therefore the same
distance $d$) if and only if they have the same closure. Keeping
this in mind, the converse of theorem 1 also holds as follows: if
$E(\psi,S)$ is an entanglement monotone, then the closure
$\overline{S}$ is a SI set. This can be shown easily using the
following observation: a state $\ket{\psi}$ is in the closure
$\overline{S}$ if and only if $E(\psi,S)=0$. Hence, if $E(\psi,S)$
is a monotone and $\ket{\psi}$ is a state in $\overline{S}$, then
by the non-negativity and monotonicity of the measure,
$E(\psi^\prime,S)=0$ for all states $\ket{\psi^\prime}$ that are
stochastically reducible from $\ket{\psi}$. This shows that all
such $\ket{\psi^\prime}$ are also in $\overline{S}$ and hence
$\overline{S}$ is SI. In conclusion, SI sets are the most general
sets that can be used in the definition (\ref{eq:E_defined}) in
order to make $E$ a monotone.

Let us also make a few remarks on SI sets. Note that the union of
two SI sets is also SI. Moreover, if $S_1$ and $S_2$ are SI, then
the monotone associated with the union $S_1\cup S_2$ is given by
\begin{equation}
  E(\psi,S_1\cup S_2) = \min(E(\psi,S_1),E(\psi,S_2))\quad.
\label{eq:E_for_union}
\end{equation}
This basically follows from the fact that the optimum state in
$\overline{S_1\cup S_2}$ which is closest to $\ket{\psi}$ [i.e.,
the state that optimizes Eqs.~(\ref{eq:d_defined}) and
(\ref{eq:E_defined})] is equal to one of the corresponding optimum
states in $\overline{S}_1$ and $\overline{S}_2$. Because of
Eq.~(\ref{eq:E_for_union}), $E(\psi,S_1\cup S_2)$ is less useful
in applications since any entanglement transformation allowed by
the monotones of $S_1$ and $S_2$ is also allowed by the monotone
of their union. Hence, when computing geometric distances, it is
sufficient to consider only ``minimal'' SI sets, which are sets
that cannot be expressed as the union of two SI sets that do not
contain each other. There is a one-to-one relation between these
minimal SI sets and the SLOCC classes. A minimal SI set
essentially contains one SLOCC-class $C$ at the top and includes
only the classes that can be stochastically reduced from $C$. Note
also that if $S_1\subset S_2$, then
\begin{equation}
  E(\psi,S_1)\geq E(\psi,S_2)\quad.
\end{equation}
The set of product states $S_P$, is contained in all SI sets, and
therefore $E(\psi,S_P)$ is the largest monotone among the
monotones investigated here.

\section{Geometric Measures for Bipartite Entanglement}
\label{sec:Bipartite}

For bipartite entanglement, only the Schmidt rank of the states,
i.e., the number of terms in the Schmidt decomposition, is
sufficient for describing the SLOCC classes and SI sets. For an
integer $n\geq1$, the SI set $S_n$ is composed of states having
Schmidt rank at most $n$. Hence, $S_1$ is the set of product
states and we have the inclusion chain
\begin{equation}
  S_1 \subset S_2 \subset \cdots \subset S_n\subset S_{n+1}\subset \cdots ~~.
\end{equation}
These are all possible SI sets for the bipartite case.

If $\ket{\psi}$ has the Schmidt rank $n$, then only $E(\psi,S_k)$
for $k=1,\ldots,n-1$ are nonzero; and
$E(\psi,S_n)=E(\psi,S_{n+1})=\cdots=0$. Let $\ket{\psi}$ have the
Schmidt decomposition
\begin{equation}
  \ket{\psi}=\sum_{i=1}^n \sqrt{\lambda_i}  \ket{i}_1\otimes\ket{i}_2\quad,
\end{equation}
and let $\lambda^\downarrow_i$ denote the Schmidt coefficients
arranged in decreasing order, i.e., $\lambda^\downarrow_1 \geq
\lambda^\downarrow_2\geq \cdots\geq \lambda^\downarrow_n$. Then,
it will be shown below that
\begin{equation}
  E(\psi,S_k) =  1-\left(\lambda^\downarrow_1+\cdots+\lambda^\downarrow_k\right)
\label{eq:E_for_bipartite}
\end{equation}
for any $k\leq n$.

In order to show Eq.~(\ref{eq:E_for_bipartite}), the following
inequality will be used. If $A$ and $B$ are arbitrary $n\times n$
square matrices, then
\begin{equation}
  \abs{\mathrm{tr}\, AB} \leq \mathbf{s}^\downarrow(A)\cdot \mathbf{s}^\downarrow(B)
     =\sum_{i=1}^n  s^\downarrow_i(A)s^\downarrow_i(B)\quad,
\label{ineq:with_singular}
\end{equation}
where $\mathbf{s}(A)$ represents the vector of singular values of
the matrix $A$ (i.e., $s_i(A)$ is the $i$th eigenvalue of
$\sqrt{A^\dagger A}$) and similarly for $B$. This inequality can
be deduced from the corresponding inequality for hermitian
matrices: if $A^\prime$ and $B^\prime$ are $n\times n$
\emph{hermitian} matrices, then
\begin{equation}
  \mathrm{tr}\,A^\prime B^\prime \leq  \boldsymbol{\lambda}^\downarrow(A^\prime)\cdot \boldsymbol{\lambda}^\downarrow(B^\prime)
    =\sum_{i=1}^n \lambda^\downarrow_i(A^\prime)\lambda^\downarrow_i(B^\prime)\quad,
\label{ineq:with_eigenvalues}
\end{equation}
where $\boldsymbol{\lambda}(A^\prime)$ and
$\boldsymbol{\lambda}(B^\prime)$ represent the vector of
eigenvalues of $A^\prime$ and $B^\prime$
respectively\cite{Bhatia}. The inequality
(\ref{ineq:with_singular}) can be proved easily by using
(\ref{ineq:with_eigenvalues}) as follows. Let
\begin{equation}
  A^\prime=\left[ \begin{array}{cc} 0 & A \\ A^\dagger & 0  \end{array}\right]\quad,\quad
  B^\prime=\left[ \begin{array}{cc} 0 & e^{i\theta} B^\dagger  \\ e^{-i\theta}B & 0  \end{array}\right]\quad.
\end{equation}
Note that $A^\prime$ is a hermitian matrix having eigenvalues $\pm
s_i(A)$; similarly for the matrix $B^\prime$. Using the inequality
(\ref{ineq:with_eigenvalues}) we get
\begin{equation}
  2\mathrm{Re}\left(e^{-i\theta}\mathrm{tr}\,AB\right) \leq 2 \mathbf{s}^\downarrow(A)\cdot \mathbf{s}^\downarrow(B)\quad.
\end{equation}
Choosing $\theta$ to be the phase angle of $\mathrm{tr}\,AB$
produces the desired inequality (\ref{ineq:with_singular}).

Now, consider any state $\ket{\phi}$ having Schmidt rank at most
$k$, i.e., $\ket{\phi}$ is any state in the SI set $S_k$. Such a
state can be expressed as
\begin{equation}
  \ket{\phi} = \frac{1}{\sqrt{\mathrm{tr}\,A^\dagger A}}
  \sum_{i,j=1}^n A_{ij} \ket{i}_1 \ket{j}_2\quad,
\end{equation}
where $A$ is any $n\times n$ matrix having matrix rank at most $k$. Hence,
\begin{equation}
  \braket{\psi}{\phi} = \frac{\mathrm{tr}\, A \sqrt{\Lambda}}{\sqrt{\mathrm{tr}\,A^\dagger A}}
\end{equation}
where $\Lambda$ represents the diagonal matrix formed from the
Schmidt coefficients of $\ket{\psi}$, i.e.,
$\Lambda_{ij}=\lambda_i \delta_{ij}$. Now, using the inequality
(\ref{ineq:with_singular}) we get
\begin{eqnarray}
  \abs{\braket{\psi}{\phi}} &\leq& \frac{\mathbf{s}^\downarrow(A)\cdot \mathbf{s}^\downarrow(\sqrt{\Lambda})}{\sqrt{\mathrm{tr}\,A^\dagger  A}}\\
    &=& \frac{\sum_{i=1}^k s^\downarrow_i(A) \sqrt{\lambda^\downarrow_i}}{\sqrt{\sum_{i=1}^k s^\downarrow_i(A)^2}}\\
    &\leq& \sqrt{\sum_{i=1}^k \lambda^\downarrow_i}
\end{eqnarray}
where we have used the fact that $A$ has at most $k$ non-zero
singular values and then invoked the Schwarz inequality in the
last step. This places an upper bound on the inner product
$\abs{\braket{\psi}{\phi}}$ for all $\ket{\phi}$ in $S_k$. This
bound is tight and can be reached by the following state in $S_k$,
\begin{equation}
  \ket{\phi_{\mathrm{max}}} = \frac{\sum_{i=1}^k \sqrt{\lambda^\downarrow_i}\ket{i^\prime}_1\otimes\ket{i^\prime}_2}{\sqrt{\sum_{i=1}^k \lambda^\downarrow_i}} \quad,
\end{equation}
where $i^\prime$ denotes the state label that has the $i$th
largest Schmidt coefficient (i.e.,
$\lambda^\downarrow_i=\lambda_{i^\prime}$). This shows that
$E(\psi,S_k)$ is given by Eq.~(\ref{eq:E_for_bipartite}).

The LOCC transformation rules for bipartite pure states have been
determined for both deterministic\cite{Nielsen99} and
probabilistic\cite{Jonathan99} transformations. It should be noted
that the necessary and sufficient conditions for the possibility
of a given entanglement transformation are exactly the condition
that all of the geometric distance monotones are non-decreasing.
In other words, an initial state $\ket{\psi}$ can be converted
into the final states $\ket{\psi_i}$ with respective probabilities
$p_i$, if and only if
\begin{equation}
  E(\psi,S_k) \geq \sum_i p_i  E(\psi_i,S_k)\quad\textrm{for~all}~k\quad.
\end{equation}
Hence, LOCC transformation rules for the bipartite states can be
expressed in terms of monotones which have a simple geometric
meaning. Note that a similar, but different, projection-operator
based expression has been given for these monotones by Barnum and
Linden\cite{Barnum01}.

\section{Geometric Measures for Three Qubits}
\label{sec:ThreeQubits}

The simplest multipartite configuration is the three qubits which
has been extensively investigated. There are six SLOCC classes for
three qubits: the multipartite classes GHZ and W; the classes for
bipartite entanglement A-BC, AB-C, C-AB; and the completely
separable states A-B-C where A, B and C denote local
parties\cite{Dur00}. The stochastic reducibility relation between
these classes are as follows: Any bipartite entangled state can be
obtained by SLOCC operations either from GHZ class or W class and
a product state can be obtained from all the other classes.

\begin{figure}
\centering
\includegraphics[scale=0.36]{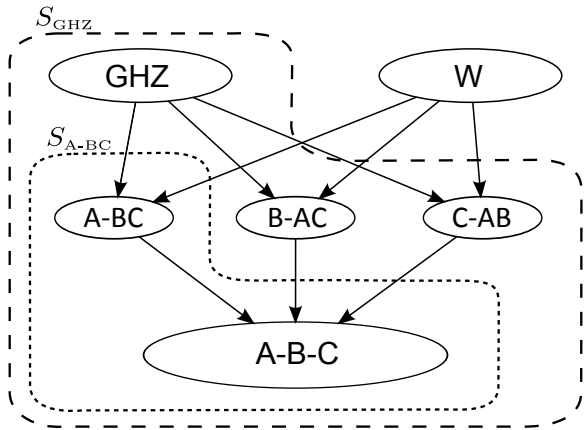}
\caption{SI sets generated by the GHZ and the A-BC classes}
\label{fig:SI}
\end{figure}

This hierarchy allows us to construct ten non-trivial SI sets. Six
of those are minimal, containing a top SLOCC class and all the
other classes that can be obtained from it. Hence, as shown in
Fig. \ref{fig:SI}
\begin{eqnarray}
S_{\textrm{GHZ}} &=&\textrm{GHZ}\cup\textrm{A-BC}\cup\textrm{AB-C}\cup\textrm{C-AB}\cup\textrm{A-B-C}\nonumber \\
S_{\textrm{A-BC}} &=& \textrm{A-BC}\cup\textrm{A-B-C}\nonumber
\end{eqnarray}
and similarly for the other classes. There are four non-trivial
unions of any two or all three of $S_{A-BC}$, $S_{AB-C}$ and
$S_{B-AC}$, which are non-minimal SI sets; but these are not worth
investigating for discovering new useful monotones.

Among the minimal SI sets, the product states $S_{A-B-C}$ produce
the original geometric distance that has been investigated before.
The SI sets generated by bipartite entangled states, i.e.,
$S_{A-BC}$, $S_{AB-C}$ and $S_{B-AC}$, give rise to monotones that
can be computed analytically by the expression derived in section
\ref{sec:Bipartite}. The SI set $S_{GHZ}$ is dense in the complete
set of states. In other words, all states outside $S_{GHZ}$,
namely the states in the W class, are the limit of some sequence
of states in $S_{GHZ}$. In fact, the generic W-class state
\begin{equation}
  \ket{\psi} =\ket{\beta_1\alpha_2\alpha_3}+\ket{\alpha_1\beta_2\alpha_3}+\ket{\alpha_1\alpha_2\beta_3}
\end{equation}
is the limit of
$\ket{\phi(\epsilon)}=(\ket{\gamma_1^\epsilon\gamma_2^\epsilon\gamma_3^\epsilon}-\ket{\alpha_1\alpha_2\alpha_3})/\epsilon$
as $\epsilon$ tends to zero, where
$\ket{\gamma_i^\epsilon}=\ket{\alpha_i}+\epsilon\ket{\beta_i}$ for
$i=1,2,3$. For this reason, the geometric distance and the
associated monotone for this SI set is identically zero, i.e.,
\begin{equation}
  d(\psi,S_{GHZ})=E(\psi,S_{GHZ})=0\quad,
\end{equation}
for all $\ket{\psi}$.

Finally, the SI set $S_W$ generated by the W class gives rise to a
new monotone. This monotone is non-zero only for states in the GHZ
class. It may appear that $E(\psi,S_W)$ is similar to the
three-tangle\cite{Coffman00}, which is also zero on $S_W$ and
non-zero only for the GHZ states. However, it turns out that
$E(\psi,S_W)$ and the three-tangle are independently useful as it
will be argued below.

First, note that if $e_1,\cdots,e_n$ are entanglement monotones
and $f(t_1,t_2,\ldots,t_n)$ is a concave function which is
increasing in each argument $t_i$, then $e^\prime$ defined by
\begin{equation}
  e^\prime(\psi)=f(e_1(\psi),e_2(\psi),\ldots,e_n(\psi))\quad,
\end{equation}
is also a monotone\cite{Gingrich02}. In this case, we will say
that the new monotone $e^\prime$ can be generated from
$e_1,e_2,\ldots,e_n$. For applications, $e^\prime$ has no use
whatsoever (if all $e_i$ can be computed) since any entanglement
transformation allowed by all $e_i$ is also allowed by $e^\prime$.
It might be of interest to investigate which of the known
monotones can be generated from the others. By using only a few
numerical evidences, it is possible to show that a given set of
monotones cannot be generated from each other. This can be done
either by finding an example against the increasing property or an
example against the concavity property of the function $f$.

For the current case of three qubits, there are five nontrivial
monotones based on the geometric distance, namely
$E(\psi,S_{A-B-C})$, $E(\psi,S_{A-BC})$, $E(\psi,S_{AB-C})$,
$E(\psi,S_{B-AC})$ and $E(\psi,S_W)$, and another monotone, the
three-tangle $\tau$. It can be shown that none of these are
generated from the others. Numerical calculations carried out
indicate that for any of these six functions, it is possible to
find a pair of states where the transformation between them is
allowed by the other five functions while forbidden by the
selected function. This means that all of these monotones are
independently useful in analyzing entanglement transformations.

The following is a simple example that shows that $E(\psi,S_W)$
cannot be generated from the other geometric measures and the
three tangle. Let
\begin{eqnarray}
    \ket{\psi} &=& \frac{1}{\sqrt{10}}(3\ket{000}+\ket{111})~~, \\
    \ket{\varphi} &=&    \frac{1}{\sqrt{N}}\left(\ket{000}-\ket{\beta\beta\beta}\right)~~,
\end{eqnarray}
where $\ket{\beta}=(\ket{0}+2\ket{1})/\sqrt{5}$ and $N$ is the
normalization factor. The function $E(\cdot,S_W)$ indicates a
transformation ordering different than those of the tangle and the
other monotones as shown in Table \ref{table:one}. This shows that
$E(\cdot,S_W)$ is a new entanglement monotone.

\begin{table}[h!]
\begin{tabular}{|c|ccc|}
        \hline
                          & $\ket{\psi}$ & & $\ket{\varphi}$ \\
        \hline
        $E(\cdot,S_{A-B-C})$  & $0.1$ & $<$ & $0.5143$ \\
        $E(\cdot,S_{AB-C}) $  & $0.1$ & $<$ & $0.3643$ \\
        $E(\cdot,S_{AC-B}) $  & $0.1$ & $<$ & $0.3643$ \\
        $E(\cdot,S_{A-BC}) $  & $0.1$ & $<$ & $0.3643$ \\
        $E(\cdot,S_{W})    $  & $0.09$& $>$ & $0.0464$ \\
        $\tau$                & $0.36$& $<$ & $0.6175$ \\
        \hline
\end{tabular}
\caption{The values of the selected entanglement measures for two
states that cannot be converted into each other. The geometric
monotones are computed numerically by an iterative algorithm that
converge to local extrema. Algorithm is repeated for several
initial random configuration for finding the global extremum. This
example shows that the monotonicity of $E(\psi,S_W)$ does not
follow from the other measures shown in the table.}
\label{table:one}
\end{table}

Another point that must be mentioned is, in contrast with the
bipartite case, the insufficiency of the geometric measures alone
for deciding the possibility of a given entanglement
transformation. As an example consider the following state in GHZ
class,
\begin{equation}
\ket{\Phi(z;c_1,c_2,c_3)}=\frac{1}{\sqrt{N}}\left(\ket{000}+z\ket{\beta_1\beta_2\beta_3}\right)~~,
\end{equation}
where $\ket{\beta_i}=c_i\ket{0}+\sqrt{1-c_i^2}\ket{1}$, $c_i$ are
real with $0\leq c_i<1$ and $z$ is a complex number. It is obvious
that the states $\ket{\Phi(z;c_1,c_2,c_3)}$ and
$\ket{\Phi(z^*;c_1,c_2,c_3)}$, where $z^*$ represents the complex
conjugate, have the same values for the three-tangle and all
geometric monotones. In a recent study\cite{Turgut09}, the rules
for deterministic entanglement transformations between
multipartite states with tensor rank 2 have been established.
According to these rules, when none of $c_i$ are zero and $z$ is
neither real nor on the unit circle, then these two states cannot
be converted into each other. Moreover, if it is possible to
transform $\ket{\Phi(z;c_1,c_2,c_3)}$ to some GHZ state
$\ket{\psi}$, then it is not possible to convert
$\ket{\Phi(z^*;c_1,c_2,c_3)}$ into $\ket{\psi}$. This example
clearly shows that the geometric monotones and the three tangle
are not sufficient for deciding on the possibility of
transformations. Hence, there must be another monotone that is not
derivable from all of these, which changes value under the complex
conjugation of the $z$ parameter.

\section{Conclusion}
\label{sec:Conclusion}

In this article, a more general approach is taken to the geometric
measure of entanglement in pure states by replacing the set of
product states with a set which is invariant under stochastic
reducibility relation. In this way, a number of new entanglement
monotones can be obtained. Moreover, it is argued that these
measures exhaust all pure-state monotones whose definition are
based on the geometric distance to a set, since the closure of
such sets must be SI. Consequently, these monotones contain
previous generalizations\cite{Barnum01,Blasone08} of the
geometric measure.

These measures quantify not only the amount, but also the flavor
of entanglement where by flavor we mean the type of entanglement
associated with each SLOCC class. The original geometric measure
$E(\psi,S_P)$, where $S_P$ is the set of product states,
quantifies the property of being entangled, meaning that the state
$\ket{\psi}$ is unentangled if and only if this measure vanishes.
In contrast to this, $E(\psi,S)$ for non-product SI sets $S$
essentially quantifies the difference of the flavor of entanglement
in $\ket{\psi}$ from the flavor associated with $S$. In other words,
an $S$ based characterization of entanglement is obtained. We have
$E(\psi,S)=0$ if and only if either $\ket{\psi}$ has an identical
flavor with the states in $S$ or otherwise it can be well
approximated with those states with desirably high fidelity.

\begin{acknowledgments}
K.U. acknowledges the financial support of The Scientific and
Technical Research Council of Turkey (T\"UB\.ITAK).
\end{acknowledgments}


\begin{thebibliography}{99}

\bibitem{Horodecki09} R. Horodecki, P. Horodecki, M. Horodecki, and K. Horodecki, Rev. Mod. Phys. \textbf{81}, 865 (2009).

\bibitem{Plenio07} M. B. Plenio and S. Virmani, Quant. Inf. Comp. \textbf{7}, 1 (2007); arXiv:quant-ph/0504163.

\bibitem{Bennett00} C. H. Bennett, S. Popescu, D. Rohrlich, J. A. Smolin, and A. V. Thapliyal, Phys. Rev. A \textbf{63}, 012307 (2000).

\bibitem{Dur00} W. D\"{u}r, G. Vidal, and J. I. Cirac, Phys. Rev. A \textbf{62}, 062314 (2000).

\bibitem{Wei03} T.-C. Wei and P. M. Goldbart, Phys. Rev. A \textbf{68}, 042307 (2003).

\bibitem{Shimony95} A. Shimony, Fundamental Problems in Quantum Theory, edited by D. M. Greenberger and A. Zelinger (1995), vol. 755 of New York Academy Sciences Annals, pp. 675-679.

\bibitem{Barnum01} H. Barnum and N. Linden, J. Phys.: Math. and Gen. \textbf{34}, 6787 (2001).

\bibitem{Vidal00} G. Vidal, J. Mod. Opt. \textbf{47}, 355 (2000).

\bibitem{Biham01} O. Biham, M. A. Nielsen, T. A. Osborne Phys. Rev. A \textbf{65}, 062312 (2001).

\bibitem{Grover97} L. K. Grover, Phys. Rev. Lett. \textbf{79}, 325 (1997).

\bibitem{Hayashi06} M. Hayashi, D. Markham, M. Murao, M. Owari and S. Virmani, Phys. Rev. Lett. \textbf{96}, 040501 (2006).

\bibitem{Gross09} D. Gross, S. T. Flammia and J. Eisert, Phys. Rev. A \textbf{102}, 190501 (2009).

\bibitem{Wei05} T.-C. Wei, D. Das, S. Mukhopadyay, S. Vishveshwara, P. M. Goldbart, Paul M. Phys. Rev. A \textbf{71}, 060305 (2005).

\bibitem{Orus08} R. Or\'us, Phys. Rev. Lett. \textbf{100}, 130502 (2008).

\bibitem{Walther05} P. Walther, K. J. Resch, and A. Zeilinger, Phys. Rev. Lett. \textbf{94} 240501 (2005).

\bibitem{Blasone08} M. Blasone, F. Dell\char39{}Anno, S. De Siena and F. Illuminati, Phys. Rev. A, \textbf{77}, 062304 (2008).

\bibitem{Hughston93} L. P. Hughston, R. Jozsa, and W. K. Wootters, Phys. Lett. A \textbf{183}, 14 (1993).

\bibitem{Bhatia} R. Bhatia, \textit{Matrix Analysis} (Springer-Verlag, New York, 1997).

\bibitem{Nielsen99} M. A. Nielsen, Phys. Rev. Lett. \textbf{83}, 436 (1999).

\bibitem{Jonathan99} D. Jonathan and M. B. Plenio, Phys. Rev. Lett. \textbf{83}, 1455 (1999).

\bibitem{Coffman00} V. Coffman, J. Kundu and W. K. Wootters, Phys. Rev. A \textbf{61}, 052306 (2000).

\bibitem{Gingrich02} R. M. Gingrich, Phys. Rev. A \textbf{65}, 052302 (2002).

\bibitem{Turgut09} S. Turgut, Y. G\"{u}l, and N. K. Pak, arXiv:0907.3960 (2009).

\end{thebibliography}
\end{document}